\documentclass{article}
\usepackage[utf8]{inputenc}
\usepackage{amsmath, amssymb}
\usepackage{braket}
\usepackage{tikz}
\usepackage{eucal}
\usepackage[hidelinks]{hyperref}
\usepackage{cite}
\usepackage{bbm}

\makeatletter
\newsavebox{\@brx}
\newcommand{\llangle}[1][]{\savebox{\@brx}{\(\m@th{#1\langle}\)}%
  \mathopen{\copy\@brx\kern-0.5\wd\@brx\usebox{\@brx}}}
\newcommand{\rrangle}[1][]{\savebox{\@brx}{\(\m@th{#1\rangle}\)}%
  \mathclose{\copy\@brx\kern-0.51\wd\@brx\usebox{\@brx}}}
\makeatother

\usepackage{titling}
\usepackage{authblk}
\title{Infrared-safe scattering without photon vacuum transitions and time-dependent decoherence}
\author{Dominik Neuenfeld\thanks{dneuenfe@phas.ubc.ca}}
\affil{Department of Physics and Astronomy, University of British Columbia, \mbox{Vancouver, BC V6T 1Z1 Canada}}

\begin{document}
\maketitle
\begin{abstract}
Scattering in $3+1$-dimensional QED is believed to give rise to transitions between different photon vacua. We show that these transitions can be removed by taking into account off-shell modes which correspond to Li\'enard-Wiechert fields of asymptotic states. This makes it possible to formulate scattering in $3+1$-dimensional QED on a Hilbert space which furnishes a single representation of the canonical commutation relations (CCR). Different QED selection sectors correspond to inequivalent representations of the photon CCR and are stable under the action of an IR finite, unitary S-matrix. Infrared divergences are cancelled by IR radiation. Using this formalism, we discuss the time-dependence of decoherence and phases of out-going density matrix elements in the presence of classical currents. The results demonstrate that although no information about a scattering process is stored in strictly zero-energy modes of the photon field, entanglement between charged matter and low energy modes increases over time.
\end{abstract}
\newpage
\section{Introduction}
Theories with long range forces suffer from IR divergences which set all S-matrix elements between Fock space states to zero. The reason is that any non-trivial scattering process produces an infinite number of low energy quanta of radiation and states which contain an infinite number of excitations are not Fock space states. However, in the calculation of scattering probabilities, one can sum over all possible additional emissions of soft quanta in incoming and outgoing states to obtain finite, inclusive quantities \cite{Bloch1937,Yennie1961,Kinoshita:1962ur,Lee:1964is,Weinberg1965} and theories with IR divergences can still be tested to high precision. As shown in \cite{Carney2017}, the construction of inclusive quantities yields an essentially completely decohered outgoing density matrix in the momentum basis and thus in this formulation, the description of scattering processes is inherently non-unitary. It should, however, be noted that this is not a flaw of the theory, but rather a flaw in our Fock space description and we should expect that there exists a better way of formulating scattering.

At last for Abelian gauge theories, the \emph{dressed formalisms} devised in \cite{Chung1965,Kibble1968a,Kulish1970,Bagan2000ChargesConstruction}  remove the IR divergences by including the radiation as coherent states in incoming and/or outgoing states.\footnote{The work of \cite{Bagan2000ChargesConstruction,Bagan2000ChargesRenormalisation} does not use dressed states, but equivalently expands fluctuations of the field around classical backgrounds which depend on the momenta of the particles involved.} This is called \emph{dressing} the in- and out-states. S-matrix elements between dressed states are finite and there is no need to calculate inclusive quantities. However, due to the infinite number of soft-modes, the dressed states are not Fock space states. Instead, as we will discuss in section \ref{sec:representations}, they live in representations of the photon canonical commutation relations (CCR) which are different from the standard Fock representation. Physically speaking, one could either say that states in different CCR representations differ by an infinite number of low-energy excitations or that they represent states which are expanded around classical backgrounds which differ at arbitrarily long wavelengths. Since the radiation produced in scattering depends on the momenta of incoming and outgoing charges, a dressed state which contains a charged particle with momentum $\mathbf p$ will be in a different CCR representation than a dressed state containing a charged particle with momentum $\mathbf q \neq \mathbf p$. In particular, this means that the associated photon vacuum states are not related by a unitary transformation. Thus, scattering states generally have different photon vacua and one says that scattering induces vacuum transitions \cite{Kapec2017}.

However, the fact that generic out-states consist of superpositions of states in different CCR representations becomes an issue if one wants to ask questions about the information content or the dynamics of low energy modes, since a meaningful comparison of the photon content between different states in different representations is impossible. A related problem recently mentioned in \cite{Gomez:2018war} is that the entirety of dressed states is non-separable \cite{Kulish1970}, i.e.~they do not have a countable basis, and thus existing dressed formalisms do not allow for the definition of a trace. And in fact, when using an IR cutoff to make the trace over IR modes well-defined, the reduced density matrix of the hard modes again essentially complete decoheres once the cutoff is removed \cite{Carney2017a}.

The soft photon production which is responsible the vacuum transitions is well approximated by a classical process, but a classical analysis suggests the number of zero-modes should stay constant: although the radiation fields which are classically produced during scattering modify the vector potential at arbitrarily long wavelengths, this change is compensated by the change of the Li\'enard-Wiechert potentials sourced by the charges. Hence, taking the off-shell modes of the classical field into account, the dynamics of the zero-modes become completely trivial and no vacuum transitions should happen.

In this paper we will argue that for quantum electrodynamics this picture is accurate even at the quantum level. We develop a new dressed formalism for QED in which the asymptotic Hilbert spaces carry only a single representation of the canonical commutation relations. In other words, all relevant photon states only differ by a finite amount of excited modes. Moreover, the representations for in- and out-states are unitarily equivalent. This implies that the S-matrix is a manifestly unitary operator. Our proposal is a modification of the dressed state formalism of \cite{Kulish1970}. In addition to coherent states describing radiation, we also incorporates classical electric fields into the definition of states and approximate the time-evolution at late times. The outgoing density matrix of any scattering is IR finite and tracing-out IR modes of the field is well-defined and does not completely decohere the density matrix at finite times. This allows for an IR safe investigation of scattering at late but finite times and enables us to discuss information theoretic properties of quantum states, e.g.~time evolution of entanglement.

\begin{figure}[t]
    \centering
    \begin{tikzpicture}
    \shade[right color=gray!50, left color=white] (-2,0) rectangle (-4,4);
    \shade[left color=gray!50, right color=white] (2,0) rectangle (4,4);
    \draw [thick] (-2,0) -- (-2,4);
    \draw [dashed] (-2.2,0) -- (-2.2,4) node[above] {\hspace{2mm}$\mathcal H^{\text{in}}$};
    \draw [thick] (2,0) -- (2,4);
    \draw [dashed] (2.2,0) -- (2.2,4) node[above] {\hspace{5mm}$\mathcal H^{\text{out}}$};
    
    \draw [->](-4.5,-0.3) -- (4.5,-0.3);
    \draw (-2.2,-0.2) -- (-2.2,-0.4) node[below] {$t_i$};
    \draw (2.2,-0.2) -- (2.2,-0.4) node[below] {$t_f$};
    
    \draw [->](-2.2,3) -- node [above] {$\mathcal T e^{-i \int_{t_i}^{-\infty} dt H_{as}(t)}$} (-4.8,3);
    \draw [->](-4.8,2) -- node [above] {$S = \mathcal T e^{-i \int_{-\infty}^\infty dt H}$}(4.8,2);
    \draw [->](4.8,1) -- node [above] {$\mathcal T e^{-i \int_{\infty}^{t_f} dt H_{as}(t)}$} (2.2,1);

    \draw (0,-0.3) node [below] {scattering region};
    \draw (-2.3,-0.3) node [below left] {asymptotic in-region};
    \draw (2.3,-0.3) node [below right] {asymptotic out-region};
    \end{tikzpicture}
    \caption[The dressed S-matrix]{The asymptotic Hilbert spaces $\mathcal H^{\text{in/out}}$ are defined at finite times $t_i$ and $t_f$. We assume the particles to be well-separated before and after $t_i$ and $t_f$, respectively (shaded regions). The time evolution of theories with long range forces is not given by the free Hamiltonian $H_0$, but approximated by the asymptotic Hamiltonian $H_\text{as}$ which takes the coupling to very low wavelength modes of the gauge field into account. Charged eigenstates of the free Hamiltonian are replaced by states dressed by transverse off-shell photons which reproduce the correct Li\'enard-Wiechert potential at long wavelengths. The dressed S-matrix $\mathbb S$ evolves a state from $t=t_i$ to $t = -\infty$ which removes the off-shell modes. It is then evolved by the standard S-matrix $S$ to $t = \infty$ and mapped onto $\mathcal H^\text{out}$ by another asymptotic time-evolution, dressing it with the correct Li\'enard-Wiechert modes. The states $\mathcal H^{\text{in/out}}$ are related by a unitary transformation.}
    \label{fig:idea}
\end{figure}
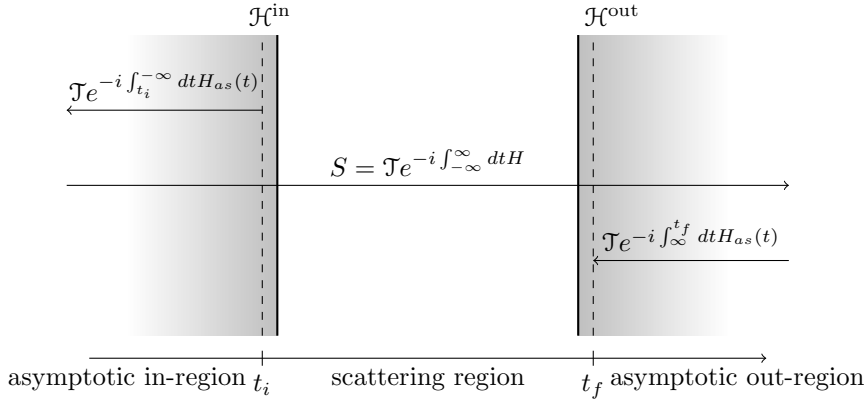

\subsection{Summary of results}
At times earlier than some initial time $t_i$ or later than some final time $t_f$, well separated states of the full theory are well approximated by states in an asymptotic Hilbert space. The dynamics relevant at long wavelengths are captured by time-evolution with an asymptotic Hamiltonian, which differs from the free Hamiltonian. This is summarized in figure \ref{fig:idea}. The asymptotic Hilbert spaces of QED are of the form
\begin{align}
\mathcal H^{\text{in}/\text{out}} = \mathcal H_m \otimes \mathcal H_\otimes(f_\lambda),
\end{align}
where $\mathcal H_m$ is the free fermion Fock space and $\mathcal H_\otimes(f_\lambda)$ is an incomplete direct product space (IDPS) (which despite the name is a Hilbert space and in particular complete) with a single representation of the photon canonical commutation relations. The precise definition is discussed in section \ref{sec:construction}. The choice of representation (equivalently, the choice of IDPS or photon vacuum) depends on a function $f_\lambda$, which generally is different for different incoming particles. $\mathcal H_\otimes(f_\lambda)$ can be understood as the image of Fock space under a (only formally defined) coherent state operator and the function $f_\lambda$ as specifying the low energy modes of the classical background.
States on this Hilbert space are dressed and take the form
\begin{align}
\| \mathbf p, \mathbf k \rrangle_{\{f_\lambda\}} = \ket {\mathbf p} \otimes W[\tilde f_\lambda(\mathbf p,\dots)] \ket {\mathbf k},
\end{align}
where $W[f_\lambda]$ are operator valued functionals which create coherent states of transverse modes whose wavefunction is given by $f_\lambda$ with polarization $\lambda$. The constraint on $\tilde f_\lambda$ is that for small photon momenta it agrees with $f_\lambda$ appearing in the definition of the photon Hilbert space.\footnote{Note that, unlike in \cite{Kulish1970}, the IR profile of soft modes in the state $\| \mathbf p, \mathbf k \rrangle_\alpha$ does not depend on $\mathbf p$ but only on $\alpha$.} This guarantees that it is a state in $\mathcal H_\otimes(f_\lambda)$. The coherent state generally contains transverse off-shell photons which ensure that at low energies, the expectation value of the photon field agrees with the classical expectation value. It also contains on-shell radiation which makes sure that the bosonic part of the dressed state lives in $\mathcal H_\otimes(f_\lambda)$. The dressed S-matrix is defined as
\begin{align}
\mathbb S = \Big(\mathcal T e^{-i \int^{t_f}_{\infty} dt H_{as}(t)} \Big) \; S\; \Big(\mathcal T e^{-i \int_{\infty}^{t_i} dt H_{as}(t)} \Big)^\dagger
\end{align}
and is a unitary operator on $\mathcal H_\otimes(f_\lambda)$ for any $f_\lambda$. The additional terms in the definition of the S-matrix remove off-shell modes from the states. This leaves states dressed with on-shell photons which are scattered by the standard S-matrix, similar to the proposal of \cite{Kulish1970}.

This framework can be used to investigate the correlation between charged particles and IR modes. Each $\mathcal H_\otimes(f_\lambda)$ inherits the trace operation from Fock space. Tracing the density matrix of a superposition of dressed states over soft modes with wavelengths above some scale $\Lambda$ yields time-dependent decoherence in the momentum eigenbasis. At late times, off-diagonal density matrix elements are proportional to
\begin{align}
  \rho^\text{reduced}_\text{off-diagonal} \propto (t \Lambda)^{-A_1} e^{A_2(t,\Lambda)}.
\end{align}
The precise form of the exponents is discussed around equation \eqref{eq:example_dampening_factor}.
The exponents are proportional to a dimensionless coupling and depend on the relative velocities of the charged matter. The factor $A_1$ is the same one found in \cite{Weinberg1965} and whose role for decoherence was discussed in \cite{Carney2017}. The dependence on time and energy scale has been found in \cite{Gomez:2018war} through a heuristic argument. The new factor $A_2$ suppresses decoherence relative to $(t \Lambda)^{-A_1}$. The only information stored in the zero-momentum modes is the information about the CCR representation and decoherence is caused by modes with non-zero momentum. As time passes, these modes become strongly entangled with the hard charges.

\subsection{Structure of the paper}
We follow the conventions of \cite{Srednicki:1019751}. QED is quantized in Coulomb gauge, since this makes the physical interpretation of our construction more obvious. Section \ref{sec:representations} reviews the construction of different representations of the CCR which are important for our purposes. Section \ref{sec:asymptotic_time_evolution} derives the asymptotic Hamiltonian and the dressed S-matrix in Coulomb gauge. The construction of the asymptotic Hilbert space is explained in section \ref{sec:construction}. Section \ref{sec:unitarity_of_s_matrix} contains a proof of the unitarity of the S-matrix. In section \ref{sec:example_classical_current} we explicitly calculate the S-matrix in the presence of a classical current and investigate the correlation between IR modes and charged particles. The density matrix of superpositions of the fields of classical currents, reduced over IR modes, decoheres with time. The conclusions comment on further directions.

\section{Representations of the canonical commutation relations}
\label{sec:representations}
\subsection{Inequivalent CCR representations}
\label{sec:inequivalent_ccr}
Theories with massless particles allow for different representations of the CCR algebra which are not unitarily equivalent. This can be easily seen in a toy model \cite{Schroer1963}. Consider the Hamiltonian
\begin{align}
    H = \int \frac{d^3\mathbf k}{(2\pi)^3 2 |\mathbf k|} |\mathbf k| a^\dagger(\mathbf k) a(\mathbf k) - \int \frac{d^3\mathbf k}{(2\pi)^3 2 |\mathbf k|} j(\mathbf k, t) (a^\dagger(\mathbf k) + a(-\mathbf k)),
\end{align}
where $j(x)$ is a real source. The Hamiltonian can be diagonalized using a canonical transformation
\begin{align}
    \label{eq:canonical_transform}
    a(\mathbf k) \to b(\mathbf k) = a(\mathbf k) + \frac{j(\mathbf k)}{|\mathbf{k}|} && a^\dagger(\mathbf k) \to b^\dagger(\mathbf k) =  a^\dagger(\mathbf k) + \frac{j^*(\mathbf k)}{|\mathbf{k}|},
\end{align}
so that the commutation relations agree for $b(\mathbf k), b^\dagger(\mathbf k)$ and $a(\mathbf k), a^\dagger(\mathbf k)$. The diagonalized Hamiltonian is given by
\begin{align}
    \label{eq:canonical_transform_hamiltonian}
    \tilde H = \int \frac{d^3\mathbf k}{(2\pi)^3 2|\mathbf k|} |\mathbf k| b^\dagger(\mathbf k) b(\mathbf k) + \frac 1 2 \int \frac{d^3\mathbf k}{(2\pi)^3} \frac{|j(\mathbf k)|^2}{|\mathbf{k}|^2}.
\end{align}
We will assume that $\lim_{|\mathbf k| \to 0} j(\mathbf k) = \mathcal O(1)$. In this case and with appropriate falloff conditions at large momenta, $\tilde H$ is bounded from below. We will assume this in the following. The formally unitary transformation which implements the transformation in equation \eqref{eq:canonical_transform} takes the form
\begin{align}
\label{eq:bad_unitary}
    W \equiv e^F = \exp \left( \int \frac{d^3\mathbf k}{(2\pi)^3 2|\mathbf k|} \left( \frac{j(-\mathbf k)}{|\mathbf{k}|} a^\dagger(\mathbf k) - h.c. \right) \right).
\end{align}
However, $W$ is not a good operator on the representation of the $a(\mathbf k),a^\dagger(\mathbf k)$ CCR, since for example
\begin{align}
    \| F \ket 0 \|^2 = \int \frac{d^3\mathbf k}{(2\pi)^3 2 |\mathbf k|^3}|j(\mathbf k)|^2 = \infty.
\end{align}
Therefore, $W$ can merely be a \emph{formally} unitary operator. This argument shows that generally, representations of the CCR of a massless field in $3+1$ dimensions coupled to different currents will be unitarily inequivalent, which is exactly the problem we discussed in the introduction. The choice of representation of the commutation relations of the photon field will generally depend on the presence of charged particles. Before we discuss how to deal with this in the case of QED, we first need to develop some formalism.

\subsection{Von Neumann space}
\label{sec:von_neumann_space}
Formally unitary operators like the one in $\eqref{eq:bad_unitary}$ can be given a meaning as operators on a complete direct product space \cite{vonNeumann1939}, henceforth \emph{von Neumann space} $\mathcal H_\otimes$. The non-separable von Neumann space splits into an infinite number of separable \emph{incomplete direct product spaces} (IDPS) on each of which one can define an irreducible representation of the canonical commutation relations \cite{klauder1966direct}. Let us review this construction in this and the next subsection.

Given a countably infinite set of separable Hilbert spaces $\mathcal H_n$, we define the infinite tensor product space $\mathcal H'_\otimes$ as
\begin{align}
    \mathcal H'_\otimes \equiv \bigotimes_n \mathcal H_n.
\end{align}
Vectors $\ket \psi \in \mathcal H'_\otimes$ of this space are product vectors built from sequences $\ket {\psi_n}$ of normalized vectors in $\mathcal H_n$,
\begin{align}
    \ket \phi = \bigotimes_n \ket {\psi_n}.
\end{align}
Two such vectors are called equivalent, $\ket \psi \sim \ket \phi$, if and only if
\begin{align}
    \sum_n | 1 - \braket{\psi_n | \phi_n}| < \infty.
\end{align}
If the vectors are equivalent their inner product is defined via
\begin{align}
    \braket{\psi|\phi} = \prod_n \braket{\psi_n | \phi_n}.
\end{align}
If two vectors are inequivalent, their inner product is set to zero by definition. The von Neumann space $\mathcal H_\otimes$ is then defined as the space obtained by extending the definition to all finite linear combinations of the vectors in $\mathcal H'_\otimes$ and subsequent completion of the resulting space. In order to make the inner product definite, we also require that two states are equal if their difference has zero inner product with any state in $\mathcal H_\otimes$. The so-obtained space is non-separable, but splits into separable Hilbert spaces $\mathcal H_\otimes (\psi)$ called incomplete direct product spaces (IDPS). $\mathcal H_\otimes (\psi)$ consists of all vectors equivalent to $\ket \psi$.

Given a unitary operator $\mathcal U_n$ on each $\mathcal H_n$ we can define a unitary operator $\mathcal U_\otimes$ on $\mathcal H_\otimes$ through
\begin{align}
    \mathcal U_\otimes \bigotimes_n \ket{\psi_n} \equiv \bigotimes_n \mathcal U_n \ket{\psi_n}
\end{align}
and extend its definition to all states in $\mathcal H_\otimes$ by linearity. Clearly, this is not the set of all possible unitary operators on $\mathcal H_\otimes$. Multiplication and inverse of such operators is defined through multiplication and inverse of the $\mathcal U_n$. It can then be shown that these unitary operators map different IDPS onto each other, i.e.~$\mathcal U_\otimes \mathcal H_\otimes(\psi) \sim \mathcal H_\otimes(\psi')$ with $\mathcal U_\otimes \ket \psi = \ket{\psi'}$. An operator $\mathcal U_\otimes$ is a unitary operator on $\mathcal H_\otimes (\psi)$ if $\mathcal U_\otimes \ket \psi \sim \ket \psi$.

In a quantum mechanical Hilbert space physical states are only identified with vectors up to a phase. In order to make this precise in a von Neumann space we define a generalized phase. Given a set of real numbers $\lambda = \{\lambda_1, \lambda_2, \dots \}$ we define the generalized phase operator $\mathcal V_\otimes(\lambda)$ as a unitary operator with $\mathcal V_n = e^{i \lambda_n}$. If $\sum_n \lambda_n$ converges absolutely, $\mathcal V_\otimes(\lambda) = e^{i \sum_n \lambda_n}$. Two vectors which differ by a generalized phase represent the same physical state. States are called weakly equivalent  $\ket \psi \sim_w \ket \phi$, if and only if
\begin{align}
    \mathcal V_\otimes(\lambda) \ket{\psi} \sim \ket{\phi}.
\end{align}

\subsection{Unitarily inequivalent representations on IDPS}
\label{sec:unitarily_inequivalent_reps}

Given the notion of a unitary operator on a von Neumann space, we can find representations of the photon CCR \cite{Kibble1968a}. Let us define the Hilbert space $\mathcal H_\gamma$ of photon wavefunctions $f_\lambda(\mathbf k)$ which obey  
\begin{align}
   \sum_\lambda \int \frac{d^3 \mathbf k}{(2\pi)^32 |\mathbf k|} |f_\lambda(\mathbf k)|^2 < \infty.
\end{align}
The inner product is given by
\begin{align}
\label{eq:inner_product}
   \braket{g|f} = \sum_\lambda \int \frac{d^3 \mathbf k}{(2\pi)^3 2 |\mathbf k|} g^*_\lambda(\mathbf k)f_\lambda(\mathbf k).
\end{align}
We are only interested in a special class of CCR representations discussed in \cite{Kibble1968a}. We define the coherent state operator\footnote{To make contact with the previous definition in terms of modes $n$, we need to expand $f_\lambda$ in a basis $e_n$ of the space of wavefunctions and define $a_n \sim \int d^3\mathbf k e_n(\mathbf k) a_\lambda(\mathbf k)$ to be the annihilation operator on $\mathcal H_n$. }
\begin{align}
\label{eq:def_coherent_state}
W[f_\lambda] \equiv \exp\left(\int  \frac{d^3 \mathbf k}{(2 \pi)^3 2 |\mathbf k|}  \left[\sum_\lambda f_\lambda(\dots, \mathbf k, t) a^\dagger_\lambda (\mathbf k) - h.c. \right] \right)
\end{align}
which formally obeys
\begin{align}
\label{eq:formal_weyl_ccr}
    W[f_\lambda]W[g_\lambda] = \exp \left(\int \frac{d^3 \mathbf k}{(2 \pi)^3 2 |\mathbf k|} \left( g^*_\lambda f_\lambda - f^*_\lambda g_\lambda \right) \right) W[g_\lambda]W[f_\lambda].
\end{align}
By functionally differentiating this equation with respect to $f_\lambda$ and $g^*_\lambda$ at $f_\lambda = g^*_\lambda = 0$ we see that the operators $a^\dagger_\lambda(\mathbf k)$ and $a_\lambda(\mathbf k)$ obey the standard CCR. If $f_\lambda,g_\lambda$ are in elements of $\mathcal H_\gamma$ the integrals in equation \eqref{eq:formal_weyl_ccr} converge and we obtain a representation on $\mathcal H_\otimes(0)$ which consists of all states equivalent to the photon vacuum $\ket 0 = \bigotimes_n \ket{0_n}$. This is the standard Fock representation. It is clear that any operator of the form $W[h_\lambda]$ with $h_\lambda \in \mathcal H_\gamma$ is a unitary operator on Fock space.

To obtain other representations we need to find operators which obey equation \eqref{eq:formal_weyl_ccr} on an IDPS $\mathcal H_\otimes(\psi)$ which is not weakly equivalent to Fock space $\mathcal H_\otimes(0)$. (It was shown in \cite{klauder1966direct} that commutation relation representations on weakly equivalent IDPS are unitarily equivalent.) Consider the space of functions $\mathcal A_\gamma$ defined by 
\begin{align}
\label{eq:definition_a}
   \sum_\lambda \int \frac{d^3 \mathbf k}{(2\pi)^3 2 |\mathbf k| } \frac{1}{|\mathbf k|}|f_\lambda(\mathbf k)|^2 < \infty.
\end{align}
Functions which obey this inequality are still dense in $\mathcal H_\gamma$. The dual vector space $\mathcal A^*_\gamma$, taken with respect to the inner product, equation \eqref{eq:inner_product}, consists of functions for which
\begin{align}
\label{eq:definition_astar}
   \sum_\lambda \int \frac{d^3 \mathbf k}{(2\pi)^3 2 |\mathbf k| } \frac{|\mathbf k|}{|\mathbf k|+1}|f_\lambda(\mathbf k)|^2 < \infty
\end{align}
and $\braket{g|f}$ is well defined for all $g \in \mathcal A_\gamma^*$ and $f \in \mathcal A_\gamma$.
Let us define the state $\ket {h} = W[h_\lambda]\ket 0$, where $h_\lambda$ lies in $\mathcal A^*_\gamma$, but not in $\mathcal A_\gamma$. Since $W[h_\lambda]$ formally diverges, the state $\ket h$ is inequivalent to the photon vacuum $\ket 0$ (even weakly). This time, operators $W[f_\lambda]$ with $f_\lambda \in \mathcal H_\gamma$ do not yield a representation of the CCR on $\mathcal H_\otimes(h)$, since
\begin{align}
\begin{split}
    \bra h W[f_\lambda]\ket h = &\exp\left( - \frac 1 2 \int \frac{d^3 \mathbf k}{(2 \pi)^3 2 |\mathbf k|} |f_\lambda|^2\right) \exp \left(\int \frac{d^3 \mathbf k}{(2 \pi)^3 2 |\mathbf k|} \left( h^*_\lambda f_\lambda - f^*_\lambda h_\lambda \right) \right)
    \end{split}
\end{align}
and the integral in the argument of the second exponential will generally diverge.
However, if we choose $f_\lambda \in \mathcal A_\gamma$, the phase converges and we obtain a representation, this time on the separable space $\mathcal H_\otimes(h)$ which can be obtained from Fock space by the formally unitary operator $W[h]$. These are the representations we will need in the following.

\section{Asymptotic time-evolution and definition of the S-matrix}
\label{sec:asymptotic_time_evolution}
\subsection{The naive S-matrix}
In the standard treatment of scattering in quantum field theory, one defines the S-matrix essentially as
\begin{align}
\label{eq:s_matrix_definition_limit}
S_{\beta,\alpha} \simeq \lim_{t' / t'' \to \mp \infty} \bra{\beta} e^{-iH(t'' - t')} \ket{\alpha}.
\end{align}
However, already in free theory it is clear that the limits $t' \to - \infty$ and $t'' \to \infty$ do not exist due to the oscillating phase at large times. More carefully we take the states $\ket{\alpha}_\text{in}/ \ket{\beta}_\text{out}$ at some fixed times $t_{i/f}$ and define the S-matrix as 
\begin{align}
\label{eq:Smatrix_standard}
S_{\beta,\alpha} = \lim_{t' / t'' \to \mp \infty} \bra{\beta}_\text{out} e^{iH_0(t''-t_f)} e^{-iH(t'' - t')} e^{-iH_0(t'-t_i)}\ket{\alpha}_\text{in}.
\end{align}
$H_0$ is the free Hamiltonian in which the mass parameter takes its physical value. At times later (earlier) than $t_f$ ($t_i$) we assume that all particles are well separated such that their time-evolution can approximately be described by the free Hamiltonian. The contribution to phase factors coming from the renormalized Hamiltonian $H = H_0 + H_{\text{int}}$ cancels the one coming from the free evolution as $t',t'' \to \mp \infty$. We can remove the dependence on $t_{i/f}$ by redefining the S-matrix $S \to e^{iH_0(t_f - t_i)} S$.\footnote{Oftentimes one chooses the convention that $t_f = t_i = T$, i.e.~the incoming and outgoing particles are defined on the same, arbitrary timeslice.} Going to the interaction picture the S-matrix can then be brought into the form
\begin{align}
\label{eq:smatrix_textbook}
S = \mathcal T e^{-i \int_{-\infty}^\infty dt \, H_{\text{int}}(t)},
\end{align}
where the Schrödinger-picture fields in the interaction Hamiltonian $H_\text{int}$ are replaced by fields evolving with the free Hamiltonian $H_0$ which gives rise to the time-dependence.

However, it is well known that the free-field approximation is not valid for QED even at late times, since the interaction falls off too slowly. Mathematically, the problem is that the expression for the S-matrix, equation \eqref{eq:Smatrix_standard}, does not converge \cite{dollard}. Physically, the issue is that massless bosons given rise to a conserved charge (e.g. electric charge in QED or ADM mass in gravity) which can be measured at infinity as an integral over the long range fields. Turning off the coupling completely at early and late times, no field is created. In this paper we use canonically quantized QED in Coulomb gauge. One might argue that the conserved charge is already taken into account by the solution to the constraint equation, which creates a Coulomb field around the source. However, for all but stationary particles, this is not the correct field configuration. Well-separated particles with non-vanishing velocity should be accompanied by the correct Li\'enard-Wiechert field which differs from the Coulomb field by transverse off-shell modes. Again, these modes can only be excited if the coupling is not turned off completely.

\subsection{The asymptotic Hamiltonian}
In order to understand which terms of the full Hamiltonian remain important at early and late times, let us approximate how the states evolve if they do not interact strongly for a long time. We ignore all UV issues which are dealt with by using renormalization and consider the normal ordered version of the interaction Hamiltonian,
\begin{align}
H_{\text{int}} \sim - e \int d^3 \mathbf x :\bar \psi \mathbf \gamma_i \psi:(\mathbf x) \cdot \mathbf A^i(\mathbf x) + \iint d^3 \mathbf x d^3 \mathbf y \frac{:\psi^\dagger \psi (\mathbf x)\psi^\dagger \psi (\mathbf y):}{4\pi|\mathbf x - \mathbf y|}.
\end{align}
In the asymptotic regions it is then assumed that the fields, masses and couplings take their physical values instead of the bare ones. In \cite{Kulish1970} it was shown that at late times coupling to long-wavelength photon modes still remain important. Here we will take a slightly different route to arrive at the exact same expression for the \emph{asymptotic Hamiltonian}, i.e.~the Hamiltonian which approximates time evolution at very early and late times.

The normal ordered current in the interaction picture in momentum space is given by
\begin{align}
\begin{split}
:j^\mu(\mathbf x): \sim e \sum_{s,t} \iint \frac{d^3 \mathbf p d^3 \mathbf q}{(2 \pi)^6 4 E_\mathbf{p} E_\mathbf{q}} \left( b_s^\dagger(\mathbf p) b_t(\mathbf q) \overline u_s(\mathbf p) \gamma^\mu u_t(\mathbf q) e^{-i (p-q) x} \right. \\ \left. -  d_t^\dagger(\mathbf q) d_s(\mathbf p) \overline v_s(\mathbf p) \gamma^\mu v_t(\mathbf q) e^{i (p-q) x} + \dots \right),
\end{split}
\end{align}
where we have omitted terms proportional to $b_s^\dagger(\mathbf p) d_t^\dagger(\mathbf q)$ and $b_t(\mathbf q) d_s(\mathbf p)$. They correspond to pair creation or annihilation with the emission or absorption of a high energetic photons. In the asymptotic regions it should be a reasonable assumption to ignore these effects. Generally, we do not want external momenta to strongly couple to the current. Thus we restrict the integral over $\mathbf q$ to a small shell around $\mathbf p$ and set $\mathbf p = \mathbf q$ everywhere except in the phases. After a Fourier transform and keeping only leading order terms in $|\mathbf k|$ we obtain the \emph{asymptotic current},
\begin{align}
\begin{split}
    :j_\text{as}^\mu(\mathbf k, t): & \sim e \sum_s \int \frac{d^3 \mathbf p}{(2 \pi)^3 2 E_\mathbf{p}} \frac{p^\mu}{E_\mathbf{p}} \left( b_s^\dagger(\mathbf p) b_s(\mathbf p)  - d_s^\dagger(\mathbf p) d_s(\mathbf p) \right) e^{-i \mathbf v_\mathbf{p} \mathbf k t}\\
    & \sim e \int \frac{d^3 \mathbf p}{(2 \pi)^3 2 E_\mathbf{p}} \frac{p^\mu}{E_\mathbf{p}} \rho(\mathbf p) e^{-i \mathbf v_\mathbf p \mathbf k t},
\end{split}
\end{align}
where we have defined $\rho(\mathbf p) = \sum_s \left( b_s^\dagger(\mathbf p) b_s(\mathbf p)  - d_s^\dagger(\mathbf p) d_s(\mathbf p) \right)$ and $\mathbf v_\mathbf p = \mathbf p/E_\mathbf p$.
At late and early times, the free Hamiltonian in equation \eqref{eq:Smatrix_standard} should thus be replaced by the time-dependent \emph{asymptotic Hamiltonian},
\begin{align}
    H_{\text{as}}(t) = H_0 + V_{\text{as}}(t),
\end{align}
which is obtained by replacing the current with the asymptotic current. The interaction potential $V_\text{as}(t)$ which replaces the interaction Hamiltonian is given in the interaction picture by
\begin{align}
\label{eq:asymptotic_potential}
V_\text{as}(t) =  - \int_\text{IR} \frac {d^3 \mathbf k}{(2 \pi)^3} \left( :\mathbf j^i(- \mathbf k, t): \mathbf A^i(\mathbf k, t) - \frac 1 {2 |\mathbf k|^2} :j^0(\mathbf k, t)j^0(-\mathbf k, t):\right).
\end{align}
The domain of integration is restricted to soft modes. The first term describes the coupling of transverse photon degrees of freedom to the transverse current,
\begin{align}
\label{eq:asymptotic_potential_transverse}
 V_\text{as}^{(1)}(t) =- \int_\text{IR} \frac {d^3 \mathbf k}{(2 \pi)^3 2 |\mathbf k|} \mathbf j^i(\mathbf k, t)  \left[ \varepsilon^{* i}_{\lambda}(- \mathbf k) a_\lambda (-\mathbf k) e^{- i |\mathbf k| t } +\varepsilon^{i}_{\lambda}(\mathbf k) a_\lambda^{\dagger } (\mathbf k) e^{i |\mathbf k| t }\right],
\end{align}
with a sum over the spatial directions $i$ implied. The second term,
\begin{align}
\label{eq:asymptotic_potential_coulomb}
 V_\text{as}^{(2)}(t) = \frac{e^2}{2} \int \frac{d^3 \mathbf p}{(2 \pi)^3 2 E_\mathbf{p} } \int_\text{IR} \frac {d^3 \mathbf k}{(2 \pi)^3} \frac 1 {|\mathbf k|^2} :\rho(\mathbf p) j^0(\mathbf q,t): e^{- i \mathbf v_\mathbf p \mathbf k t},
\end{align}
gives the energy of a charge in a Coulomb field created by a second charge.

\subsection{The dressed S-matrix}
In the spirit of equation \eqref{eq:Smatrix_standard} we define the \emph{dressed S-matrix} as an operator which maps the asymptotic Hilbert space of incoming states $\mathcal H^\text{in}$ to the asymptotic Hilbert space of outgoing states $\mathcal H^\text{out}$,
\begin{align}
\label{eq:smatrix}
\mathbb S = \lim_{t' / t'' \to \mp \infty} \mathcal T e^{-i \int_{t''}^{t_f}dt H_\text{as}(t)} e^{-iH(t'' - t')} \mathcal T e^{-i \int_{t_i}^{t'}dt H_\text{as}(t)},
\end{align}
where $\mathcal T$ denotes time-ordering. It seems plausible that in the case of QED this expression has improved convergence over equation \eqref{eq:s_matrix_definition_limit}, since $H_\text{as}$ takes into account the asymptotic behavior of $H$.\footnote{It has been conjectured in \cite{Dybalski2017} that a similar expression in the context of the Nelson model converges. However, other work \cite{Laddha:2018myi} indicates that there might be subleading divergences coming from current-current interactions.} In order to simplify the expression for the S-matrix and relate it to the standard expression, we insert the identity, $\mathbbm 1 = e^{-i H_0 (t''-t_f)} e^{i H_0 (t''-t_f)}$ and $\mathbbm 1 = e^{-i H_0 (t'-t_i)} e^{i H_0 (t'-t_i)}$, between the time ordered exponentials and the full time evolution. We then obtain
\begin{align}
\begin{split}
\mathbb S = \lim_{t' / t'' \to \mp \infty} U(t_f, t'') \; S \; U(t',t_i),
\end{split}
\end{align}
where $S = e^{i H_0 (t''-t_f)} e^{-iH(t'' - t')} e^{-i H_0 (t'-t_i)}$ reduces to the usual S-matrix in non-dressed formalisms, equation \eqref{eq:smatrix_textbook}, once the limits are taken. The unitaries $U(t_1,t_0)$ obey the differental equation
\begin{align}
i \frac{\partial}{\partial t_1} U(t_1, t_0) = V_\text{as}(t_1) U(t_1, t_0), 
\end{align}
where $V_\text{as}$ is in the interaction picture and given by equation \eqref{eq:asymptotic_potential}.
The solution to this is standard\footnote{See, e.g.~chapter 4.2 of \cite{Peskin:1995ev}.}
\begin{align}
U(t_1, t_0) = \mathcal T e^{-i\int_{t_0}^{t_1}dt V_\text{as}(t)}.
\end{align}
We can bring this into an even more convenient form \cite{Kulish1970} by splitting $U(t_1,t_0)$ in the following way, 
\begin{align}
\begin{split}
U(t_i, t_0) &= \mathcal T e^{-i \left( \int_{t_i-\epsilon}^{t_i} + \dots + \int_{t_0}^{t_0+\epsilon} \right) dt V_\text{as}(t)} \\
&= \mathcal T e^{-i  \int_{t_i-\epsilon}^{t_i} dt V_\text{as}(t)} \dots e^{-i\int_{t_0}^{t_0+\epsilon}  dt V_\text{as}(t)} \\
&= \mathcal T e^{-i  \int_{t_i-\epsilon}^{t_i}dt V_\text{as}(t)} \dots \mathcal T e^{-i\int_{t_0}^{t_0+\epsilon}  dt V_\text{as}(t)}.
\end{split}
\end{align}In the limit $\epsilon \to 0$ we can remove the time-ordering symbols. Since $[V_\text{as}(t),V_\text{as}(t')]$ only depends on $\rho(\mathbf p)$ which commutes with all operators we can use the Baker-Campbell-Hausdorff formula $e^A e^B = e^{A+B} e^{1/2[A,B]}$ to combine the exponentals into
\begin{align}
\label{eq:field_dressing}
U(t_i, t_0) = e^{-i \int_{t_0}^{t_i} dt V_\text{as}(t)} e^{- \frac 1 2 \int_{t_0}^{t_i} dt \int_{t_0}^t dt' [V_\text{as}(t), V_\text{as}(t')]}.
\end{align}
The first factor couples currents to the transverse electromagnetic potential and also contains the charge-charge interaction given in equation \eqref{eq:asymptotic_potential_coulomb}. The second factor makes sure that $U(t_2, t_1)U(t_1, t_0) = U(t_2, t_0)$. We are interested in the limit where $t_0 \to -\infty$. In this case the second factor can be calculated as follows. Since the density $\rho(\mathbf p)$ commutes with all operators present in the asymptotic potential, the only relevant contributions to the commutator come from the photon annihilation and creation operators. The unequal-time commutator of the asymptotic potential with itself is given by 
\begin{align}
[V_\text{as}(t),V_\text{as}(t')] = \int_\text{IR} \frac {d^3 \mathbf k}{(2 \pi)^3 2 |\mathbf k|}  \mathbf j_\text{as}^\perp(-\mathbf k, t) \mathbf j_\text{as}^\perp(\mathbf k, t')\left(e^{i|\mathbf k|(t'-t)} - e^{-i|\mathbf k|(t-t')}\right),
\end{align}
with the transverse current $\mathbf j^{\perp,i}(\mathbf k, t) = \sum_{\lambda} \varepsilon^{i*}_\lambda(\mathbf k)\varepsilon^{j}_\lambda(\mathbf k) \mathbf j^{j}(\mathbf k, t)$.
We can now perform the integral over $t'$ and drop the boundary conditions as $t = -\infty$ knowing that in any final calculation they will be canceled by the corresponding term coming from the full Hamiltonian. The result is
\begin{align}
\begin{split}
H_c^\perp(t) = - &\frac 1 2 \int_{-\infty}^t dt'[V(t), V(t')] \\
= \frac{i}{2}&  \int_\text{IR} \frac {d^3 \mathbf k}{(2 \pi)^3 2 |\mathbf k|} \int \frac {d^3 \mathbf p}{(2 \pi)^3 2 E_\mathbf{p}} \frac{\mathbf v_\mathbf p - \frac{\mathbf k(\mathbf k \cdot \mathbf v_\mathbf p)}{|\mathbf k|^2}}{|\mathbf k|- \mathbf k \cdot \mathbf v_\mathbf p } \\& \qquad \times \Big[ :\rho(\mathbf p) \mathbf j_\text{as}(-\mathbf k, t): e^{- i \mathbf k \mathbf v_\mathbf p t} + h.c. \Big],
\end{split}
\end{align}
where we have used that $\rho(p):j_\text{as}(-\mathbf k, t): = :\rho(p) j_\text{as}(-\mathbf k, t):$ up to terms that are renormalized away \cite{Kulish1970}.
This corrects the phase due to the Coulomb energy, equation \eqref{eq:asymptotic_potential_coulomb}, to
\begin{align}
    e^{i \Phi(t)} \equiv e^{i \int_{-\infty}^t dt' (H_c(t') + H_c^\perp(t'))},
\end{align}
which gives the phase due to the energy of a charge in the Li\'enard-Wiechert field of another charge. The total asymptotic time evolution takes the form
\begin{align}
\label{eq:in_dressing}
U(-\infty,t_i) =  e^{i \Phi(t)} e^{i \int_{-\infty}^{t} dt' V^{(1)}_\text{as}(t')} .
\end{align}
An analogous expression follows for $U(t_f,\infty)$, where we have to drop the boundary terms at $t = \infty$.

\section{Construction of the asymptotic Hilbert space}
\label{sec:construction}

\subsection{The asymptotic Hilbert space}
\label{sec:the_asymptotic_hilbert_space}
We can finally discuss the asymptotic Hilbert space. For now, we will ignore free photons and moreover focus on a single particle. The generalization to many particles and the inclusion of free photons is straight forward and will be done later.
We require that our asymptotic states evolve with the asymptotic Hamiltonian instead of the free one. Naively, we might be tempted to think that our asymptotic particle agrees with a free field excitation at some time $t$. However, as discussed in the previous section, if our field couples to a massless boson this will generally not be correct. Given a charged excitation of momentum $\mathbf p$ we define
\begin{align}
\label{eq:state_def_1}
\begin{split}
\| \mathbf p \rrangle^\text{in}_{\mathbf p}  &\equiv U(t_i,-\infty)(\ket{\mathbf p}^\text{in} \otimes \ket 0)\\
&\equiv \ket{\mathbf p}^\text{in} \otimes W[f^\text{in}_\lambda(\mathbf p, \mathbf k,t)] \ket 0.
\end{split}
\end{align}
The state $\ket{\mathbf p}^\text{in}$ is a free field fermion Fock space state defined at time $t_i$ and $\ket 0$ is the photon Fock space vacuum. $U(t_i,-\infty)$ was given in equation \eqref{eq:in_dressing} and does not change the matter component of the state. We can therefore write its action as an operator on the photon Hilbert space, $W[f^\text{in}_\lambda]$, with $W[\; \cdot \;]$ given in equation \eqref{eq:def_coherent_state}. In \eqref{eq:in_dressing}, we have dropped the boundary term at $-\infty$. This is analogous to the standard procedure one uses to get the electric field of a current at a time $t$ from the retarded correlator. The subscript in equation \eqref{eq:state_def_1} indicates that the asymptotic Hilbert space containing the state $\| \mathbf p \rrangle_\mathbf{p}^\text{in}$ accordingly is 
\begin{align}
\label{eq:asymptotic_hilbert_space}
    \mathcal H_{\text{as}} = \mathcal H_\text{m} \otimes \mathcal H_\otimes(f^\text{in}_\lambda(\mathbf p, \mathbf k, t_i)),
\end{align}
where $\mathcal H_\text{m}$ is the standard free fermion Fock space and $H_\otimes(f^\text{in}_\lambda(\mathbf p, \mathbf k, t_i))$ is an incomplete direct product space which carries a representation of the canonical commutation relations for the photon as explained in the previous subsection. Performing the integral in $U(t_i,-\infty)$, we can determine $f^\text{in}_\lambda(\mathbf p, \mathbf k, t_i)$ to be
\begin{align}
f^\text{in}_\lambda(\mathbf p, \mathbf k, t) = - e \frac{p \cdot \varepsilon_\lambda(\mathbf k)}{p \cdot k} \theta(k^\text{max} - |\mathbf k|)  e^{- i v\cdot k t_i }.
\end{align}
Here, $p^\mu$ and $k^\mu$ are on-shell and $v^\mu = p^\mu/E_\mathbf p$. The Heaviside function makes sure that only modes with wave number smaller than $k^\text{max}$ are contained in the dressing. 
Analogously, we can construct an out-states as
\begin{align}
\label{eq:state_def_out}
\begin{split}
\| \mathbf p \rrangle^\text{out}_{\mathbf p}  &\equiv U(t_f,\infty)(\ket{\mathbf p}^\text{out} \otimes \ket 0)\\
&\equiv \ket{\mathbf p}^\text{out} \otimes W[f^\text{out}_\lambda(\mathbf p, \mathbf k,t)] \ket 0,
\end{split}
\end{align}
and
\begin{align}
f^\text{out}_\lambda(\mathbf p, \mathbf k, t_f) = - e \frac{p \cdot \varepsilon_\lambda(\mathbf k)}{p \cdot k} \theta(k^\text{max} - |\mathbf k|) e^{- i v\cdot k t_f } = f^\text{in}_\lambda(\mathbf p, \mathbf k, t_f).
\end{align}
In the following, we will leave the sum over $\lambda$ and the dependence of $f_\lambda(\mathbf p, \mathbf k, t)$ on $\mathbf k$ and $t$ implicit. It can be checked by power counting that the exponent of
\begin{align}
\bra 0 W[f^\text{in}_\lambda(\mathbf p)] \ket 0 = \exp \left({\frac 1 2 \int  \frac{d^3 \mathbf k}{(2 \pi)^3 2 |\mathbf k|} |f^\text{in}_\lambda(\mathbf p)|^2}\right) 
\end{align}
is IR divergent, so $W[f^\text{in}_\lambda(\mathbf p)]$ is not a unitary operator on Fock space. It can also be checked that $W[f^\text{in}_\lambda(\mathbf p)]$ obeys equation \eqref{eq:definition_astar} so that the commutation relation representation is inequivalent to the Fock space representation. On the other hand $W[f^\text{out}_\lambda(\mathbf p) - f^\text{in}_\lambda(\mathbf p)]$ is a unitary operator on any representation since its argument is in $\mathcal A_\gamma$, defined through equation \eqref{eq:definition_a}. This operator maps $\text{in}$-states to $\text{out}$-states and it follows that $\mathcal H_\otimes(f(\mathbf p)^\text{out}_\lambda) = \mathcal H_\otimes(f(\mathbf p)^\text{in}_\lambda)$. Since the Hilbert spaces are related by unitary time-evolution using the asymptotic Hamiltonian, in the following we will oftentimes drop the $\text{in}$ and $\text{out}$ labels on the states. Equivalently we can set $t_i = t_f = T$ without affecting any argument in the following.

The coherent state of transverse modes in equation \eqref{eq:state_def_1} which accompanies the matter field $\ket{\mathbf p }^\text{in}$ is \emph{not} a cloud of on-shell photons. The reason is that the time-dependence of $f^\text{in}_\lambda(\mathbf p)$ modifies the dispersion relation of the modes created by this coherent state from $E_{\mathbf k} = |\mathbf k|$ to $E_{\mathbf k} = \mathbf k \mathbf v$. To understand the role of these modes consider the expectation values of the four-potential in such a dressed state,
\begin{align}
\llangle {\mathbf p} \| A^0  \| \mathbf p \rrangle &= \int \frac{d^3\mathbf k}{(2 \pi)^3} \frac{1}{|\mathbf k|^2} \llangle {\mathbf p} \| j^0 (\mathbf k, t)  \| \mathbf p \rrangle e^{i \mathbf k \mathbf x},\\
\llangle {\mathbf p} \| \mathbf A \| \mathbf p \rrangle &= e \int_0^{k^\text{max}} \frac{d^3 \mathbf k}{(2\pi)^3 2 |\mathbf k|} \frac{\mathbf v_\mathbf p - \frac{\mathbf k(\mathbf k \cdot \mathbf v_\mathbf p)}{|\mathbf k|^2}}{|\mathbf k|- \mathbf k \cdot \mathbf v_\mathbf p } \left[ e^{i \mathbf k (\mathbf x - \mathbf v_\mathbf p t)} + h.c. \right] \llangle \mathbf p \| \mathbf p \rrangle.
\end{align}
The expectation value of $\mathbf A$ agrees with the classical 3-vector potential of a point charge moving in a straight line with velocity $\mathbf v_\mathbf p$ at long wavelength which passes through $\mathbf x = 0$ at $t=0$,
\begin{align}
j^\mu(\mathbf k, t) = e v^\mu e^{- i \mathbf v_\mathbf p \mathbf k t}.
\end{align}
In other words, the dressed state constructed above obeys Ehrenfest's theorem at long wavelengths. If we had not dressed the state, we would have found $\llangle \mathbf p \| \mathbf A \| \mathbf p \rrangle$, the corresponding electric field would have been only the Coulomb field of a static charge.\footnote{In the case of a plane wave the charge distribution is smeared over all of space.} 

Given two momenta $\mathbf p \neq \mathbf q$, the Hilbert spaces $\mathcal H_\otimes(f_\lambda^\text{in}(\mathbf p))$ and $\mathcal H_\otimes(f_\lambda^\text{out}(\mathbf q))$ are weakly inequivalent. To see this, note that $\tilde W \equiv W[f_\lambda^\text{out}(\mathbf q)] W^\dagger[f_\lambda^\text{in}(\mathbf p)]$ maps $\mathcal H_\otimes(f_\lambda^\text{in}(\mathbf p))$ to $\mathcal H_\otimes(f_\lambda^\text{in}(\mathbf q))$ and up to a phase equals $\tilde W = W[f_\lambda^\text{in}(\mathbf q) - f_\lambda^\text{in}(\mathbf p)]$. If the Hilbert spaces were equivalent $\tilde W$ would have to be a unitary operator on $\mathcal H_\otimes(f_\lambda^\text{in}(\mathbf p))$. However, it is easy to see that $f_\lambda^\text{in}(\mathbf q) - f_\lambda^\text{in}(\mathbf p)$ does not obey \eqref{eq:definition_a} and thus the two Hilbert spaces cannot be equivalent. Since we have started with the claim, that we want all in- and out-states to be elements of the Hilbert space \eqref{eq:asymptotic_hilbert_space}, it seems our program has failed. However, this is too naive. Assume we scatter an initial state $\| \mathbf p \rrangle$ off of a classical potential. Our outgoing state will be a superposition of different momentum eigenstates. However, the state $\| \mathbf q \rrangle_\mathbf{q}$ will not be part of this superposition. A scattering process produces an infinite number of long-wavelength photons as bremsstrahlung, but $\| \mathbf q \rrangle_\mathbf{q}$ contains no such radiation. The IR part of the classical radiation field produced during scattering from momentum $\mathbf p$ to $\mathbf q$ is created by a coherent state operator
\begin{align}
\begin{split}
R(\mathbf p,\bar{\mathbf q}) &\equiv W[f^\text{rad}_\lambda(\mathbf p, \mathbf k, t) - f^\text{rad}_\lambda(\mathbf q, \mathbf k, t)] \\
&= W[f^\text{rad}_\lambda(\mathbf p, \mathbf k, t)] W^\dagger[f^\text{rad}_\lambda(\mathbf q, \mathbf k, t)]
\end{split}
\end{align}
with
\begin{align}
f^\text{rad}_\lambda(\mathbf p, \mathbf k) = \frac{e  p \cdot \varepsilon_\lambda(\mathbf k)}{p \cdot k} g(|\mathbf k|)\approx -f^\text{in}_\lambda(\mathbf p, \mathbf k, 0).
\end{align}
The bar in the definition of $R(\mathbf p, \bar{\mathbf q})$ denotes that the terms containing $\mathbf q$ come with a relative minus sign. Here, $g(|\mathbf k|)$ is a function which goes to $1$ as $|\mathbf k| \to 0$ and can be chosen at will otherwise. Thus the state which is obtained by scattering an excitation with momentum $\mathbf p$ into an excitation with momentum $\mathbf q$ plus the long wavelength part of the corresponding bremsstrahlung is given by 
\begin{align}
\label{eq:dressed_state_general}
\| \mathbf q \rrangle_{\mathbf p} \equiv \ket {\mathbf q} \otimes W[f_\lambda^\text{in}(\mathbf q)] R(\mathbf q,\bar{\mathbf p}) \ket 0
\end{align}
up to a finite number of photons. This state contains the field of the state $\| \mathbf q \rrangle_{\mathbf q}$ as well as the radiation produced by scattering the state $\| \mathbf p \rrangle_{\mathbf p}$ to momentum $\mathbf q$ at long wavelengths.

The operator $W[f^\text{in}(\mathbf q)] R(\mathbf p,\bar{\mathbf q})$ again is not a unitary operator on any CCR representation. However, the combination 
\begin{align}
\label{eq:convergent_combination}
W[f_\lambda^\text{out}(\mathbf q)] R(\mathbf q,\bar{\mathbf p})W^\dagger[f_\lambda^\text{in}(\mathbf p)]
\end{align}
converges on Fock space. The convergence up to phase is easy to see since up to a phase, equation \eqref{eq:convergent_combination} equals $W[f_\lambda^\text{out}(\mathbf q) + f_\lambda^\text{rad}(\mathbf q) - f_\lambda^\text{in}(\mathbf p) - f_\lambda^\text{rad}(\mathbf p)]$ and since the function in the argument vanishes as $|\mathbf k| \to 0$ it clearly satisfies equation \eqref{eq:definition_a}. It is an easy exercise to prove that the phase is also convergent. We will give an example below. This shows that the states $\| \mathbf p \rrangle_\mathbf{p}$ and $\| \mathbf q \rrangle_\mathbf{p}$ live in the same subspace $\mathcal H_\otimes(f_\lambda^\text{in}(\mathbf p))$. Moreover, all states which are physically accessible from $\| \mathbf p \rrangle_\mathbf{p}$ must contain radiation. States of the form $\| \mathbf q \rrangle_\mathbf{\mathbf p}$ are constructed to precisely contain the IR tail of the classical radiation. Hence, all single fermion states which are physically accessible take the form of equation \eqref{eq:dressed_state_general} up to a finite number of photons and thus live in the same separable IDPS. With the appropriate dressing, also multi-fermion states and thus all physically accessible states live in this subspace. Note that this structure is different to existing constructions \cite{Kibble1968a,Kulish1970,Gomez:2018war}, where an out-state is generally a superposition of vectors from inequivalent subspaces of $\mathcal H_\otimes$.

\subsection{Multiple particles and classical radiation backgrounds}
The generalization to multiple particles is straight forward. Given a state which contains multiple charges with momenta $\mathbf p_1, \mathbf p_2, \dots$, the operator $U^\dagger(t_i,-\infty)$ acts on the photon state as\footnote{In the case of multiple particle species with different charges, we should replace $e \to e_i$ in the definition of $f_\lambda^\text{in}(\mathbf p)$.}
\begin{align}
    W\left[\sum_i f_\lambda^\text{in}(\mathbf p_i)\right]
\end{align}
and maps the Fock space vacuum into a a different separable Hilbert space $\mathcal H_\otimes(\sum_i f_\lambda^\text{in}(\mathbf p_i))$, which acts as our asymptotic photon Hilbert space. Similarly, we can define a coherent state operator
\begin{align}
    R(\mathbf p_1, \mathbf p_2, \dots; \overline{\mathbf q_1}, \overline{\mathbf q_2}, \dots)
\end{align}
which lets us define states 
\begin{align}
    \| \mathbf q_1, \mathbf q_2, \dots \rrangle_{\{\mathbf p_1, \mathbf p_2, \dots\}} \in \mathcal H_\otimes\left(\sum_i f_\lambda^\text{in}(\mathbf p_i)\right),
\end{align}
which contain particles with momenta $\mathbf q_1, \mathbf q_2, \dots$ and the appropriate brems\-strah\-lung produced by scattering charged particles of momenta $\{\mathbf p_1, \mathbf p_2, \dots\}$ to charged particles of momenta $\{ \mathbf q_1, \mathbf q_2, \dots \}$. Up to a finite number of additional photons all out states will be of this form.

We can also incorporate classical background radiation described by
\begin{align}
    A^0 &= 0 \\
    \mathbf A &= \int \frac{d^3 \mathbf k}{(2\pi)^3 2 |\mathbf k |} \left[ h_\lambda(\mathbf k) e^{i  k x} + h.c. \right]
\end{align}
with $\lim_{|\mathbf k| \to 0} |\mathbf k| h_\lambda(\mathbf k) = \mathcal O(1)$, i.e.~backgrounds which contain an infinite number of additional infrared photons. In the presence of charged particles with momenta $\mathbf p_1, \mathbf p_2, \dots$ the corresponding asymptotic Hilbert space is $\mathcal H_\otimes(h_\lambda + \sum_i f_\lambda^\text{in}(\mathbf p_i))$.

\subsection{Comments on the Hilbert space}
\label{sec:comments_on_hilbert_space}
The construction presented in this paper has a number of properties which are known to be realized in theories with long range forces in $3+1$ dimensions, but usually glossed over.

\paragraph{Existence of selection sectors.}
The existence of selection sectors in four-dimensional QED and gravity is well established \cite{Frohlich:1978bf,Ashtekar1987AsymptoticQuantization} and has recently been rediscovered \cite{Kapec:2015ena}. In the present construction, the choice of selection sector corresponds to a choice of representation of the canonical commutation relations on a separable Hilbert space $\mathcal H_\otimes(\psi) \subset \mathcal H_\otimes$. That these are indeed selection sectors will be shown in the next section where we prove that $\mathbb S$ is unitary.

\paragraph{Charged particles as infraparticles.}
It was shown in \cite{Schroer1963,Frohlich:1978bf,Buchholz:1986uj} that there are no states in QED (or more generally in theories with long range forces), which sit exactly on the mass-shell $p^2 = -m^2$. Our construction reproduces this behavior. Although $P\cdot P \| \mathbf p \rrangle_{\mathbf p} = -m^2 \| \mathbf p \rrangle_{\mathbf p}$, the state is not non-normalizable.\footnote{$P$ is the 4-momentum operator.} 
A normalizable state must be built from a superposition of different states $\| \mathbf q \rrangle_{\mathbf p}$. However, any other state in $\mathcal H_{\mathbf p}$ contains extra photons and thus cannot be on the mass-shell $p^2 = -m^2$. Also note that in \cite{Carney:2018ygh} it was argued that consistent scattering of wavepackets in theories with long range forces in four dimensions requires to take superpositions of particle states including photons.

\paragraph{Spontaneous breaking of Lorentz invariance.}
The spontaneous breaking of Lorentz invariance in QED has already been noted in \cite{Frohlich:1978bf,Frohlich:1979uu} (see also \cite{Balachandran:2013wsa}). In our construction, there is an infinite number of possible $\mathcal H_\otimes(\psi)$ one can choose from. This choice spontaneously breaks Lorentz invariance. The states $\| \mathbf p \rrangle_{\mathbf p}$ and $\| \mathbf q \rrangle_{\mathbf q}$ describe boosted versions of the same configuration, namely a charged particle in the absence of radiation. However, as shown above they live in inequivalent representations. Thus, a Lorentz transformation cannot be implemented as a unitary operator on $\mathcal H_{\otimes}(f^\text{in}_\lambda(\mathbf p))$. An analogous argument applies for any configuration of charged particles $\mathbf p_1, \mathbf p_2, \dots$.

\section{Unitarity of the S-matrix}
\label{sec:unitarity_of_s_matrix}
The form of the S-matrix follows from equation \eqref{eq:smatrix},
\begin{align}
\mathbb S = U(t_f, \infty)\, S \, U^\dagger(t_i,-\infty),
\end{align}
with $U(t_1,t_0)$ given in equation \eqref{eq:field_dressing}. The operator $S$ is the textbook S-matrix. Comparing to equation \eqref{eq:state_def_1} we see that the role of the operators $U(t_f, \infty)$, $U^\dagger(t_i,-\infty)$ is to remove the part of the dressing which corresponds to the classical field. Thus, the off-shell dressing $U(t_i,-\infty)$ in the definition of the asymptotic states, equation \eqref{eq:dressed_state_general}, can be ignored whenever we are calculating S-matrix elements.

Consider the action of the dressed S-matrix on {$\| \mathbf p_1, \mathbf p_2, \dots \rrangle_{\{f_\lambda\}} \in \mathcal H_\otimes(f_\lambda)$.} We establish unitarity on $\mathcal H_\otimes(f_\lambda)$ by showing that dressed S-matrix elements between states with given $f_\lambda$ are finite, as well as that dressed S-matrix elements between states of different separable subspaces, i.e. various $f_\lambda, \tilde f_\lambda$ with different IR asymptotics vanish. Unitarity then follows from unitarity of $U$ in the von Neumann space sense and unitarity of S.

For the sake of clarity we will neglect the possibility of a classical background radiation field in the following. Taking this possibility into account corresponds to acting with some coherent state operator $\tilde R$ on the Fock space vacuum and does not affect the proof. We take an otherwise arbitrary, dressed in-state
\begin{align}
\| \text{in} \rrangle & =  \ket {\mathbf p_1, \dots} \otimes W[f^\text{in}_\lambda(\mathbf p_1) + \dots ] R(\mathbf p_1,\dots; \overline{\mathbf q_1}, \dots) \ket {\mathbf k_1, \dots}
\end{align}
and similarly define a general out-state 
\begin{align}
\| \text{out} \rrangle & =  \ket {\mathbf p'_1, \dots} \otimes W[f^\text{out}_\lambda(\mathbf p'_1) + \dots ]R(\mathbf p'_1,\dots; \overline{\mathbf q_1}, \dots) \ket {\mathbf k'_1, \dots}.
\end{align}
Both states are elements of $\mathcal H_{\mathbf q_1, \dots}$.
For ease of notation, we will omit the ellipses \dots and indices in the following. The S-matrix elements take the form
\begin{align}
\begin{split}
\mathbb S_{out, in} &{}= \llangle \text{out} \| U(t_f, \infty) S U^\dagger(t_i, -\infty) \| \text{in} \rrangle \\ 
& = \Big(\bra{\mathbf p'} \otimes \bra {\mathbf k'} R^\dagger(\mathbf p;\overline{\mathbf q})\Big) \; S \; \Big( \ket {\mathbf p} \otimes R(\mathbf p; \overline{\mathbf q}) \ket {\mathbf k_1} \Big).
\end{split}
\end{align}
It was conjectured in \cite{Kapec2017} and shown in \cite{Choi:2017ylo} (see also \cite{Carney:2018ygh}) that we can move dressings through the S-matrix without jeopardizing the IR-finiteness. We can therefore move all $\mathbf q_i$ dependent terms on one side and obtain 
\begin{align}
\begin{split}
 \llangle \text{out} \| R^\dagger(\mathbf p';\overline{\mathbf q}) S R(\mathbf p; \overline{\mathbf q}) \| \text{in} \rrangle & =  \llangle \text{out} \| R({\mathbf q};\overline{\mathbf p'}) S R(\mathbf p; \overline{\mathbf q})\| \text{in} \rrangle \\
 & =  \llangle \text{out} \| R(0;\overline{\mathbf p'}) S R(\mathbf p;0)\| \text{in} \rrangle + \text{(finite)}.
 \end{split}
\end{align}
Hence, the divergence structure of the matrix element is the same as the one of 
\begin{align}
\mathbb S_{\text{out}, \text{in}} \sim \Big( \bra{\mathbf p'_1, \dots} \otimes \bra {\mathbf k'_1, \dots} R^\dagger(\mathbf p'_1, \dots;0) \Big) \; S \; \Big( \ket {\mathbf p_1, \dots} \otimes R(\mathbf p_1, \dots; 0) \ket {\mathbf k_1, \dots} \Big).
\end{align}
However, these are just Faddeev-Kulish amplitudes which are known to be IR finite \cite{Kulish1970}.

Let us now show that if $\| \mathbf p_1, \dots \rrangle_{\mathbf q_1, \dots}$ and $\| \mathbf p'_1, \dots \rrangle_{\mathbf q'_1, \dots}$ live in inequivalent representations, the matrix element vanishes. We again omit the ellipses and indices. Consider 
\begin{align}
\mathbb S_{\text{out}', \text{in}} &= \llangle \text{out} \| U(t_f, \infty) S U^\dagger(t_i, -\infty) \| \text{in} \rrangle \\ 
& = \Big( \bra{\mathbf p'} \otimes \bra {\mathbf k'} R^\dagger(\mathbf p';\overline{\mathbf q'})  \Big) \; S \; \Big( \ket {\mathbf p} \otimes R(\mathbf p; \overline{\mathbf q}) \ket {\mathbf k} \Big).
\end{align}
Moving the dressing through the S-matrix, we find that up to finite terms
\begin{align}
\mathbb S_{\text{out}', \text{in}} &\sim \bra{\text{out}'} R(\mathbf q', \overline {\mathbf q}) R^\dagger({\mathbf p'};0) S R(\mathbf p; 0) \ket{\text{in}}.
\end{align}
The previous proof showed that $R^\dagger({\mathbf p'};0) S R(\mathbf p;0)$ is a unitary operator on Fock space. Further, it can be shown that $R(\mathbf q', \overline {\mathbf q})$ vanishes on Fock space if $\mathbf q_1, \dots \neq \mathbf q'_1, \dots$ \cite{Carney2017}. Therefore we can conclude that the S-matrix element vanishes and have shown that the S-matrix is a stabilizer of the asymptotic Hilbert spaces defined in section \ref{sec:construction}.

\section{Example: Classical current}
\label{sec:example_classical_current}
\subsection{Calculation of the dressed S-matrix}
The formalism devised in the preceding sections can be used to investigate the time dependence of decoherence in scattering processes.
A simple example can be given by considering QED coupled to a classical current $j^\mu(x)$. The current enters with momentum $\mathbf p$ and at $x^\mu = x_0^\mu$ is deflected to a momentum $\mathbf p'$,
\begin{align}
\label{eq:classical_current}
\begin{split}
j^\mu(x) ={}& e \int_{0}^\infty d\tau \frac{p'^\mu}m \delta^{(4)}\left(x^\mu - x^\mu_0 -  \frac{p'^\mu}m \tau\right) 
\\ & \quad + e \int^{0}_{-\infty} d\tau \frac{p^\mu}m \delta^{(4)}\left(x^\mu - x^\mu_0 - \frac{p^\mu}m \tau\right).
\end{split}
\end{align}
We assume that initially no radiation is present and the current is carried by an infinitely heavy particle. The initial state of the transverse field excitations is not the Fock vacuum but $\| \text{in} \rrangle = W[f^\text{in}_\lambda(\mathbf p) ] \ket 0 $, which is the vacuum of the CCR representation $\mathcal H_\otimes(f^\text{in}_\lambda(\mathbf p))$. This state represents a situation in which the classical field of the current $j^\mu$ is present at wavelengths longer than the inverse mass. Since we deal with an infinitely massiv source, the integrals are taken over all of values of $\mathbf k$. The IR divergent Fock space S-matrix in the presence of a current can be calculated explicitely, see e.g. \cite{D.Bjorken1964}, and is given by
\begin{align}
S &= R(\mathbf q, \overline{\mathbf p}) = W[f_\lambda^\text{rad}(\mathbf q, \mathbf k) - f_\lambda^\text{rad}(\mathbf p, \mathbf k)].
\end{align}
According to our prescription, the dressed S-matrix is given by
\begin{align}
\mathbb S = W[f^\text{out}_\lambda(\mathbf q, \mathbf k, t_f) ] \; S \; W^\dagger[f^\text{in}_\lambda(\mathbf p, \mathbf k, t_i) ].
\end{align}
The out state is given by $\| \text{out} \rrangle = \mathbb S \, \| \text{in} \rrangle$ and contains the radiation field produced by the acceleration as well as a correction to the Coulomb field which depends on the outgoing current.
Combining everything, the dressed S-matrix becomes
\begin{align}
\mathbb S = W[f^S_\lambda(\mathbf p, \mathbf q, \mathbf k, t_i,t_f)] \exp \left(i e^2 \int \frac{d^3 \mathbf k}{(2 \pi)^3 2 |\mathbf k|} \Phi(\mathbf k, \mathbf q, \mathbf p) \right)
\end{align}
with
\begin{align}
\begin{split}
f^S_\lambda(\mathbf p, \mathbf q, \mathbf k, t) = &e \left(\frac{q \cdot \varepsilon_\lambda(\mathbf k)}{q \cdot k}(1-e^{i v_\mathbf q \cdot k t_f}) - \frac{p \cdot \varepsilon_\lambda(\mathbf k)}{p \cdot k}(1-e^{i v_\mathbf p \cdot k t_i})\right),\\
\Phi(\mathbf k, \mathbf q, \mathbf p) = &\left(\frac{\mathbf q^\perp}{q \cdot k} - \frac{\mathbf p^\perp}{p \cdot k}\right) \left(\frac{\mathbf q^\perp}{q \cdot k} \sin(v_\mathbf q \cdot k t_f) 
+ \frac{\mathbf p^\perp}{p \cdot k} \sin(v_\mathbf p \cdot k t_i) \right) \\ &+ \frac{\mathbf q^\perp}{q \cdot k} \frac{\mathbf p^\perp}{p \cdot k} \sin\left((t_f v_\mathbf q  - t_i v_\mathbf p ) \cdot k\right)
\end{split}
\end{align}
The superscripts on the momentum vectors $\mathbf p^\perp \equiv P^\perp(\mathbf {\hat k}) \mathbf p$ denote the part of $\mathbf p$ which is perpendicular to $\mathbf k$. The projection operator $P^\perp(\mathbf {\hat k})$ arises from the sum over polarizations, $P^\perp(\mathbf {\hat k}) = \sum_{\lambda = \pm} \varepsilon^*_\lambda(\mathbf k)\varepsilon_\lambda(\mathbf k)$. From here it is easy to see that as $|\mathbf k| \to 0$, $f^S_\lambda$ has no poles and $\Phi$ only goes like $|\mathbf k|^{-1}$. Therefore, $\mathbb S$ is a well defined unitary operator.

\subsection{Tracing out long-wavelength modes}
A big advantage of formulating scattering in terms of the dressed states introduced above is that it allows an IR divergence free definition of the trace operation the on asymptotic Hilbert space. The trace operation is inherited from Fock space. For example, a basis for the Hilbert space of photon excitations in $\mathcal H_\otimes(f^\text{in}_\lambda(\mathbf p))$ is given by 
\begin{align}
\label{eq:basis}
\begin{split}
W[-f^\text{rad}_\lambda(\mathbf p)] \ket 0, W[-f^\text{rad}_\lambda(\mathbf p)] \int \frac{d^3 \mathbf k}{(2\pi)^3 2 |\mathbf k|} a_{\lambda'}^\dagger(\mathbf k) \ket 0, \\
\dots, W[-f^\text{rad}_\lambda(\mathbf p)] \frac{1}{\sqrt{n!}} \left(\int \frac{d^3 \mathbf k}{(2\pi)^3 2 |\mathbf k|} a_{\lambda'}^\dagger(\mathbf k)\right)^n \ket 0
\end{split}
\end{align}
We could have chosen any other $\tilde f_\lambda(\mathbf p,\mathbf k, t)$ as long as $\lim_{\mathbf k \to 0} |\mathbf k | f^\text{in}_\lambda(\mathbf p,\mathbf k, t_i) = \lim_{\mathbf k \to 0} |\mathbf k | \tilde f_\lambda(\mathbf p,\mathbf k, t)$. For example we could have chosen $\tilde f_\lambda(\mathbf p,\mathbf k, t)= f^\text{out}_\lambda(\mathbf p,\mathbf k, t_f)$, since the trace is invariant under a change of basis.

As an example, let us consider a superposition of fields created by classical currents, i.e. the outgoing state is
\begin{align}
    \| \text{out} \rrangle = \frac 1 {\sqrt {2N} } \left(  W_{\mathbf q_1}  + W_{\mathbf q_2} \right) \ket 0,
\end{align}
where
\begin{align}
     W_{\mathbf q_i} \equiv W[f_\lambda^\text{out}(\mathbf q_i, \mathbf k, t)] \; W[f_\lambda^\text{rad}(\mathbf q_i, \mathbf k) - f_\lambda^\text{rad}(\mathbf p, \mathbf k)]
\end{align}
and $N$ is given by
\begin{align}
    N = 1 + \text{Re}\Big( \bra 0 W_{\mathbf q_1}^\dagger W_{\mathbf q_2} \ket 0\Big).
\end{align}
In order to calculate the reduced density matrix we split the dressing $W_{\mathbf q_i} = W_{\mathbf q_i}^\text{IR} +  W_{\mathbf q_i}^\text{UV}$ into a part we will trace over (IR) and the complement (UV). The ``IR'' part contains all modes with wavelength longer than some cutoff $\Lambda$, which is smaller than $k^\text{max}$. The reduced density matrix obtained by tracing over ``IR'' then becomes
\begin{align}
    \rho^\text{UV} = \frac 1 N &\Big(  W_{\mathbf q_1}^\text{UV} \ket 0 \bra 0  W_{\mathbf q_1}^{\text{UV}\dagger} +   \bra 0 W_{\mathbf q_2}^{\text{IR} \dagger} W_{\mathbf q_1}^{\text{IR}} \ket 0  \; \; W_{\mathbf q_1}^\text{UV} \ket 0 \bra 0  W_{\mathbf q_2}^{\text{UV}\dagger} \\
    &   + (\mathbf q_1 \leftrightarrow \mathbf q_2) \Big).
\end{align}
We see that the off-diagonal elements are multiplied by a factor of $\bra 0 W_{\mathbf q_2}^{\text{IR} \dagger} W_{\mathbf q_1}^{\text{IR}} \ket 0$ which is responsible for decoherence. A similar \emph{dampening factor} already appeared in \cite{Carney2017a}. There the calculation was done for Faddeev-Kulish dressed states and it was shown that the dampening factor has an IR divergence in its exponent which makes it vanish, unless $\mathbf q_1 = \mathbf q_2$. As we will see, using the dressing devised in this paper, the dampening factor is IR finite for finite times.

The magnitude of the dampening factor is simply the normal-ordering constant of $W_{\mathbf q_2}^{\text{IR} \dagger} W_{\mathbf q_1}^{\text{IR}}$ which is given by
\begin{align}
   \exp \left(-\frac 1 2 \int \frac{d^3 \mathbf k}{(2\pi)^3 2|\mathbf k|} \sum_{\lambda = \pm} |f^1_{\lambda} - f^2_{\lambda}|^2 \right)
\end{align}
with
\begin{align}
   f^i_{\lambda}(\mathbf q_i, \mathbf k, t) = e \frac{q_i \cdot \varepsilon_\lambda(\mathbf k)}{q_i \cdot k}(1 - e^{- i v\cdot k t}).
\end{align}
We can rearrange the terms proportional to $|f^i|^2$. We go to spherical polar coordinates and separate the $|\mathbf k|$ integral to find
\begin{align}
\int \frac{d^3 k}{2|\mathbf k|} \sum_{\lambda = \pm} |f^i_{\lambda}|^2 = e^2 \int d^2\Omega \frac{\mathbf q^\perp_i \mathbf q^\perp_i }{(q_i \cdot k)^2} \int_0^\Lambda\frac{d|\mathbf k|}{|\mathbf k|} \sin^2\left(|\mathbf k| \frac{(- v_i \cdot \hat k)}{2}t\right) 
\end{align}
The $|\mathbf k|$ integral can be performed and the result can be expressed in terms of logarithms and cosine integral functions $\text{Ci}(x)$.
\begin{align}
\begin{split}
    \int_0^\Lambda\frac{d|\mathbf k|}{|\mathbf k|} \sin^2&\left(|\mathbf k| \frac{(- v_i \cdot \hat k)}{2}t\right) \\
& = \frac 1 2 \left( \log(\Lambda t) + \gamma + \log(|v_i \cdot \hat k|) - \text{Ci}(\Lambda t |v_i \cdot \hat k|) \right).
\end{split}
\end{align}
Here, $\gamma$ is the Euler-Mascheroni constant. Using $\text{Ci}(x) \sim \gamma + \log(x) + \mathcal O (x^2)$ for small $x$, we see that at $\Lambda, t = 0$ the exponent vanishes.
The $|\mathbf k|$ integral for the cross-term involving $f^1_\lambda$ and $f^2_\lambda$ is only slightly more complicated and can also be performed. One finds
\begin{align}
\begin{split}
\int \frac{d^3 \mathbf k}{2|\mathbf k|} \sum_{\lambda = \pm} & \text{Re}(f^{1*}_{\lambda}f^2_\lambda) \\
& = 2 e^2 \int d^2\Omega \frac{\mathbf q_1^\perp \mathbf q_2^\perp }{(q_1 \cdot \hat k)(q_2 \cdot \hat k)} \int_0^\Lambda\frac{d|\mathbf k|}{|\mathbf k|} \sin\left(|\mathbf k| \frac{(- v_1 \cdot \hat k)}{2}t\right) \times \\ & \qquad \sin\left(|\mathbf k| \frac{(- v_2 \cdot \hat k)}{2}t\right)\cos\left(|\mathbf k| \frac{(- (v_1-v_2) \cdot \hat k)}{2}t\right) 
\end{split}
\end{align}
The integral evaluates to
\begin{align}
\begin{split}
    \frac {1}{4}\left( 2 \log(\Lambda t) + \gamma + \log(|v_1 \cdot \hat k|) + \log(|v_2 \cdot \hat k|) - \log(\Lambda t|(v_1-v_2) \cdot \hat k|) \right. \\
    \left. - \text{Ci}(\Lambda t |v_1 \cdot \hat k|) - \text{Ci}(\Lambda t |v_2 \cdot \hat k|) + \text{Ci}(\Lambda t |(v_1-v_2) \cdot \hat k|)\right).
\end{split}
\end{align}
Clearly, as $t \to 0$ the dampening factor becomes zero and no decoherence takes place. This is sensible is the example at hand, where we have assumed that the current changes direction at $t=0$. Different to the situation in \cite{Carney2017}, the density matrix is well defined even without an IR cutoff. In any real experiment we measure the field at very late times after the scattering process has happened and all wavelengths shorter than those that will be traced out had enough time to be produced, i.e.~$\Lambda t \gg 1$. In this limit, the integrals are dominated by the logarithms. Furthermore, we need to keep the term which contains $\text{Ci}(\Lambda t |(v_1-v_2) \cdot \hat k|) - \gamma$, since the cosine integral diverges as $v_1 \to v_2$ and $\hat{\mathbf k} \perp \mathbf v_1, \mathbf v_2$. 

Similarly, the phases of the off-diagonal terms in the density matrix can be calculated. Since we only have a single charge present, the Coulomb interactions $H_\text{c} + H^\perp_\text{c}$ does not contribute anything to the phase. The only contributions come from the normal ordering of the coherent state operators. After some cancellations and performing the integration over $|\mathbf k|$ we obtain
\begin{align}
    \exp\left(i \frac{e^2}{2(2\pi)^3} \int d^2\Omega \frac{\mathbf q^\perp_1 \mathbf q^\perp_2 }{(q_1 \cdot \hat k)(q_2 \cdot \hat k)} \text{Si}(\Lambda t (v_1 - v_2) \hat k) \right).
\end{align}

Thus, at late times, the dampening factor becomes
\begin{align}
\label{eq:example_dampening_factor}
\bra 0 W_{\mathbf q_2}^{\text{IR} \dagger} W_{\mathbf q_1}^{\text{IR}} \ket 0 = (\Lambda t)^{-A_1} e^{A_2(\Lambda, t)}
\end{align}
with
\begin{align}
\label{eq:decoherence_a_1}
    A_1 &= \frac{e^2}{2(2\pi)^3} \int d^2\Omega \left( \frac{\mathbf q^\perp_1}{q_1 \cdot \hat k} - \frac{\mathbf q^\perp_2}{q_2 \cdot \hat k} \right) \left( \frac{\mathbf q^\perp_1}{q_1 \cdot \hat k} - \frac{\mathbf q^\perp_2}{q_2 \cdot \hat k} \right) \\
\begin{split}
    \label{eq:decoherence_a_2}
    A_2(t,\Lambda) &= - \frac{e^2}{2(2\pi)^3} \int d^2\Omega \frac{\mathbf q^\perp_1 \mathbf q^\perp_2 }{(q_1 \cdot \hat k)(q_2 \cdot \hat k)} \Big( \text{Ci}(\Lambda t |(v_1-v_2) \cdot \hat k|)  \\
 & \qquad   - i \text{Si}(\Lambda t (v_1-v_2) \cdot \hat k) - \gamma - \log(\Lambda t|(v_1-v_2) \cdot \hat k|) \Big).
\end{split}
\end{align}
This is consistent with earlier results obtained in \cite{Calucci2003,Gomez:2018war}. The appearance of the factor $A_2$ makes the decoherence rate for particles milder than suggested by the term which only depends on $A_1$. The qualitative behavior at infinite times, however, reproduces exactly what has been found before based on calculations which only take the emitted radiation into account, namely that any reduced density matrix decoheres in the infinite time limit \cite{Carney2017}.

\section{Conclusions}
In this paper we presented a construction of an infinite class of asymptotic Hilbert spaces for QED which are stable under S-matrix scattering with a unitary, dressed S-matrix. The major improvement over existing work is that all asymptotic states live in the same separable Hilbert space with a single representation of the photon canonical commutation relations. Our construction relied on the fact that transverse IR modes of the Li\'enard-Wiechert field are included in the definition of the asymptotic states. This should be a good approximation if the included wavelengths are smaller than any other scale in the problem. The construction enables an analysis of the information content of IR modes in the late-time density matrix. As an example, we studied a density matrix which describes a superposition of the field of two classical currents. The reduced density matrix decoheres as a power law with time. The increase of decoherence with time shows that the entanglement of charged particles with infrared modes increases over time. The physical reason for the decoherence is that at times $t \sim \frac 1 \Lambda$ we can tell apart on- and off-shell modes with wavelengths larger than $\lambda \sim \frac 1 \Lambda$. Since charged matter is accompanied by a cloud of off-shell modes creating the correct momentum-dependend electric field, this allows to identify the momenta of the involved particles. It is instructive to compare this to the picture of conserved charges from large gauge transformations (LGT) (for a review see \cite{Strominger2018LecturesTheory}). There it is argued that a photon vacuum transition must happen since the soft charge generally changes during a scattering process, while the total LGT charge remains constant. However, our approach identified the photon vacuum with the total LGT charge, i.e. it takes into account off-shell excitations of the photon field associated with the hard part of the LGT charge. The increase of decoherence with time can be understood as learning to tell apart soft and hard charges as time goes on. Hence, in flat space scattering, no information is stored in the LGT charges, but in the way the charge splits between the hard and soft part.


This work leaves open some interesting questions. We have seen that near-zero energy modes decohere the outgoing density matrix in the momentum basis. Unlike in \cite{Carney2017, Carney2017a}, this decoherence happens although the scattering is fundamentally IR finite. Furthermore, the decoherence cannot be avoided by choosing an appropriate dressing, since we can only add radiation, i.e.,~on-shell modes, as additional dressing. At zero energy there is no difference between on and off-shell modes, however, at finite times those can be distinguished which leads to decoherence. This opens up the possibility that a similar mechanism at subleading order in the asymptotic current could also decohere additional quantum numbers like spin. Moreover, although we have constructed dressed states, we have not discussed how they can be obtained by an LSZ-like formalism from operators. Due to the presence of long wavelength modes of classical fields and radiation, the correct operators must be non-local. Presumably there should be an infinite family of operators, similar to the situation in \cite{Bagan2000ChargesConstruction,Bagan2000ChargesRenormalisation}, for each Hilbert space which must contain radiative modes in their definition. Filling in the details is left for future work. Lastly, an extension of the presented ideas to gravity would be desirable for a variety of reasons. While one might expect that a generalization to linearized gravity should be fairly straight forward, an extension beyond linear order will presumably more difficult. The discussion in the context of gravity could be interesting in the context of the Black Hole information paradox: We have seen that in our construction no information is stored in the zero-energy excitations. This agrees with statements made in \cite{Mirbabayi2016,Bousso:2017dny}. However, by waiting long enough, charged matter can be arbitrarily strong correlated with near zero-energy modes and those modes might store information. Tracing out the matter thus leaves one with a completely mixed density matrix of soft modes, which might be related to the ideas presented in \cite{Haco:2018ske}. The fact that ``softness'' is an observer-dependent notion might aid arguments in favor of complementarity. Clearly, more work is required to make these arguments more precise.

\section*{Acknowledgements}
I thank Dan Carney, Jason Pollack and Gordon Semenoff for discussions and comments on the manuscript. Furthermore, I thank Laurent Chaurette, Colby DeLisle, Jordan Wilson-Gerow and participants of PITP2018 at the IAS for an inspiring exchange of ideas, as well as David Wakeham for comments a draft. This work is supported by a UBC Four Year Doctoral Fellowship and I would like to acknowledge partial support by NSERC and the Simons Foundation.

\appendix

\bibliographystyle{unsrt}
\bibliography{bibliography}
\end{document}